# Excitation, detection, and electrostatic manipulation of terahertz-frequency range plasmons in a two-dimensional electron system

Jingbo Wu, Alexander S. Mayorov, Christopher D. Wood, Divyang Mistry, Lianhe Li, Wilson Muchenje, Mark C. Rosamond, Li Chen, Edmund H. Linfield, A. Giles Davies, and John E. Cunningham[*]

School of Electronic and Electrical Engineering, University of Leeds, Woodhouse Lane, Leeds LS2 9JT, United Kingdom.
*j.e.cunningham@leeds.ac.uk

## ABSTRACT

Terahertz time domain spectroscopy employing free-space radiation has frequently been used to probe the elementary excitations of low-dimensional systems. The diffraction limit blocks its use for the in-plane study of individual laterally defined nanostructures, however. Here, we demonstrate a planar terahertz-frequency plasmonic circuit in which photoconductive material is monolithically integrated with a two-dimensional electron system. Plasmons with a broad spectral range (up to ~400 GHz) are excited by injecting picosecond-duration pulses, generated and detected by a photoconductive semiconductor, into a high mobility two-dimensional electron system. Using voltage modulation of a Schottky gate overlying the two-dimensional electron system, we form a tuneable plasmonic cavity, and observe electrostatic manipulation of the plasmon resonances. Our technique offers a direct route to access the picosecond dynamics of confined transport in a broad range of lateral nanostructures.

## Introduction

Picosecond time-resolved measurements of low-dimensional semiconductors can reveal a diverse range of physical phenomena. Typically, a device containing a two-dimensional electron system (2DES) is subjected to free-space propagating terahertz (THz) radiation (100 GHz < $f$ < 5 THz), and either the transmitted THz response and/or the rectification response of the 2DES is then used to infer information about the system. Such experiments have provided information on, for example, coherent cyclotron resonance in a 2DES,[1,2] ultrastrong light-matter interactions between inter-Landau level transitions of a 2DES and the photonic modes of artificial resonators,[3,4] THz-wave modulation at room temperature,[5-7] and recently, the formation of plasmonic crystals using a gate-controlled 2DES.[8,9] The latter is particularly exciting, since the plasmonic cavity resonances in 2DESs on length scales of a few microns occur in the THz frequency range, offering the possibility of fabricating plasmonic circuits that can be used to manipulate THz signals.

Another class of experiments involves the planar integration of a 2DES into THz waveguides, which allows pulses to be either directly coupled into the system by ohmic contacts, using a flip-chip arrangement,[10] or coupled by proximity to a nearby THz waveguide, where they are exposed to and interact with the evanescent THz electric field.[11] In both these cases, electrical pulses are usually guided along a lithographically-defined, sub-wavelength transmission line structure formed on a separate substrate, before interacting with the 2DES. The in-plane nature of these techniques provides an enhanced interaction between the THz signal and the low-dimensional system relative to that achieved through free-space coupling. Such techniques have allowed ultrafast ballistic picosecond transport[10] and magnetoplasmon resonances[11] to be studied, albeit the latter with a somewhat limited time resolution (~10 ps) with respect to free-space THz time-domain spectroscopy. This is presumably because of dispersion in the lossy THz waveguide and low-dimensional coupled system.

Recently, we introduced an alternative technique in which growth-optimized LT-GaAs (providing THz-bandwidth pulse excitation and detection) and a high mobility 2DES channel are integrated in a single molecular beam epitaxy (MBE) wafer.[12] Here, we demonstrate that such integrated structures can be used to form broadband (up to ~400 GHz) on-chip plasmonic circuits capable of the in-plane excitation, detection, and electrostatic manipulation of 2D plasmons in quantum-confined 2DESs. The dynamic evolution of plasmon resonances in the gated 2DES region, controlled by an applied voltage, is recorded with a few-picosecond time resolution. Our methodology thus opens up a wide range of possible experiments in which broadband pulsed THz radiation is used to probe individual mesoscopic or nanoscale systems defined lithographically in a 2DES, rather than ensembles.

## Results

### Schematic and principle

A diagram of our THz 2D plasmonic circuit, in which the photoconductive material of LT-GaAs is monolithically integrated with 2DES, is shown in Figure 1a. The device was fabricated from an MBE wafer (Figure 1b), which comprised a layer of LT-GaAs along with a GaAs/AlGaAs heterostructure containing a 2DES (see further details in Methods). Two pairs of photoconductive (PC) switch contacts were then defined on the LT-GaAs layer, after it was subjected to a selective wet-etch to remove the 2DES and expose the underlying photoconductive LT-GaAs layer. A coplanar waveguide (CPW) guides THz pulses generated from, for example, PC switch S1, to an ohmic contact that is used to inject the picosecond pulses into the (73-µm-long) 2DES mesa. When the propagating THz pulse arrives at this ohmic contact, a portion of the pulse energy is injected into the 2DES, transmitted through the 2DES, and then exits through a second ohmic contact, before being coupled into the adjacent section of CPW. The first ohmic contact also reflects a portion of the propagating pulse. The time-resolved transmitted and reflected signals are then sampled at S3 or S2, respectively. As shown in Figure 1c, a 4.4-µm-wide metal gate was defined on the top of the 2DES mesa. A negative gate voltage ($V_g$) applied to this gate was used to deplete carriers and so tune the electron concentration ($n_s$) in the 2DES underneath. The voltage ($V_{th}$) required to deplete fully carriers underneath the gate at 4 K after illumination was ~−3.0 V (see Supplementary Note 1 and Figure S1).

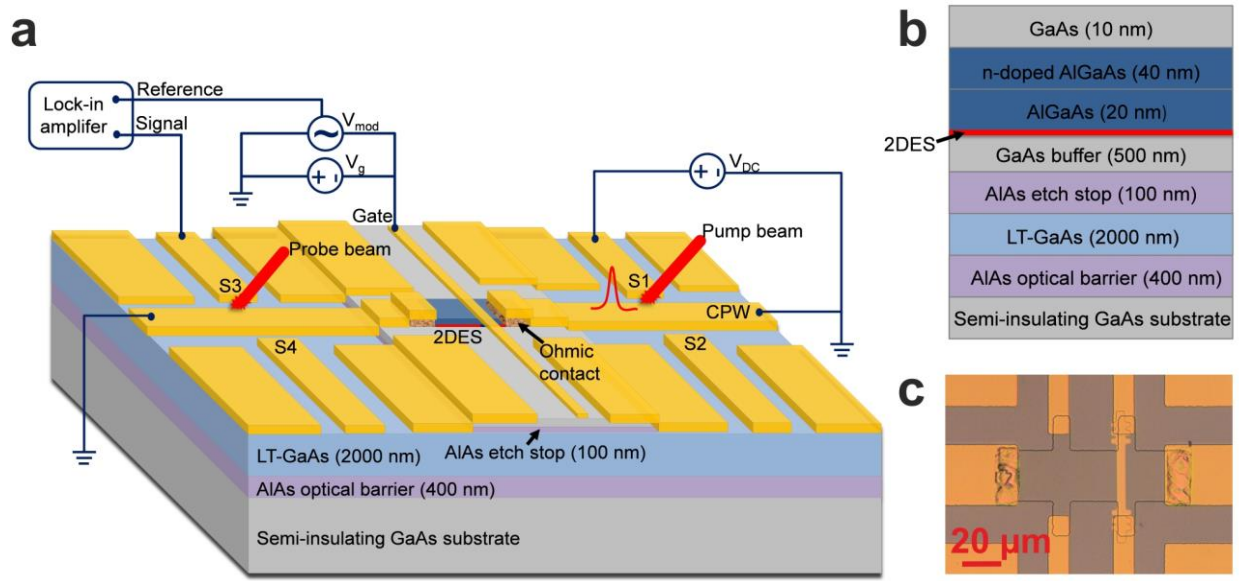

**Figure 1. Diagram of the THz 2D plasmonic circuit.** (a) Schematic diagram of the THz plasmonic circuit and the measurement arrangement for gate-modulation signals. S1/S2 and S3/S4 are two pairs of PC switches formed on opposite sides of the 2DES mesa, and are used to generate or detect the THz pulses; pulses are generated by application of a bias while under illumination by a 800 nm pulsed Ti:sapphire laser, while detection is achieved by measuring the generated photocurrent as a function of optical path time delay. (b) The layer structure of the wafer monolithically integrating the LT-GaAs and GaAs/AlGaAs heterostructure containing the 2DES (red region). (c) Microscopic image of the 2DES mesa. A 4.4-µm-long metallic gate was located on top of the 2DES mesa, and the widths of the ungated regions on either side of the gate were 19.7 µm and 48.9 µm.

Our 2DES mesa supports the propagation of 2D plasmons at THz frequencies.[13,14] Therefore, when a THz pulse is injected into the 2DES, 2D plasmons are excited over a broad frequency range, and the 2DES acts as a plasmonic transmission line. The plasmonic dispersion relation in a 2DES is given by:

$$\omega_p = \sqrt{\frac{n_s e^2}{2m^* \varepsilon_0 \varepsilon_{eff}(k)}} k, \tag{1}$$

where $e$ and $m^*$ are the charge and effective mass of electrons in GaAs, respectively, $\varepsilon_0$ is the vacuum dielectric permittivity, $\varepsilon_{eff}(k)$ is the effective relative permittivity, and $k$ is the plasmon wave number[15-17]. Unlike bulk plasmons, the resonance frequency of 2D plasmons depend on geometric size. In order to excite a resonant plasmonic mode, $k$ must satisfy the condition $k = n\pi/L$, where $n = 1, 2, 3...$, and $L$ is the length of the plasmonic cavity. The phase velocity of plasmons is then obtained using $v_p = \omega_p/k$. In the ungated regions, the 2DES behaves as a dispersive, single-wire-like plasmonic transmission line.[18] In the gated region, however, the metallic gate screens the Coulombic restoring force, which has the effect of reducing the acceleration of electrons driven by the exciting THz electric field. The plasmons in this region therefore have a much lower $v_p$ in comparison with the ungated areas. The effective permittivity $\varepsilon_{eff}(k)$ in the gated region is given by $\varepsilon_{eff}(k) = [\varepsilon_2 + \varepsilon_1 \coth(kd)]/2$, where $\varepsilon_1$ and $\varepsilon_2$ are the relative permittivity of AlGaAs and GaAs, respectively, and $d$ is the distance from the metallic gate to the 2DES.[18] The plasmon dispersion can hence be written:

$$\omega_p = \sqrt{\frac{n_s e^2}{m^* \varepsilon_0 [\varepsilon_2 + \varepsilon_1 \coth(kd)]}} k. \tag{2}$$

If the screening effect is strong (when $kd \rightarrow 0$), the transmission line effectively then acts as a parallel-plate, plasmonic waveguide, supporting a dispersionless transverse electromagnetic (TEM) mode.[8,17]

## THz pulse propagation in the 2DES

To investigate the injection into and propagation of THz pulses through the 2DES, we measured the input and transmitted pulses using what is, in effect, an on-chip THz time-domain spectrometer.[19,20] Initial measurements were performed in a continuous-flow helium cryostat at 4 K (see Supplementary Note 2 and Figure S2). Figure 2a shows the measured input and reflected pulses as a function of $V_g$. The first peak observed at 0 ps is the input pulse generated by S1, and detected by S2 through the conductive coupling of the two adjacent PC switches across the centre conductor. The second peak, centred at 9.8 ps, is the signal reflected from the ohmic contact closest to S1/S2. When a THz pulse arrives at this ohmic contact, it is partly reflected from the CPW/2DES interface, whilst a portion of the signal is injected into the 2DES mesa. As $V_g$ is reduced from 0 V to $V_{th}$ (–3.0 V), a 'shoulder' appears on the reflected time-domain signal. The amplitude of this feature increases with decreasing $V_g$, and saturates at $V_{th}$. We attribute the 'shoulder' feature to the increased reflection of the signal propagating in the 2DES by the elevated barrier under the gated region when a negative voltage is applied. When $V_g = 0$ V, and since $v_p$ in gated region is lower than that in ungated region, the mismatch of $k$ $(k = \omega/v_p)$ leads to the formation of a barrier. As $V_g$ decreases, $n_s$ and the corresponding $v_p$ in the gated region also both decrease. As a result, the mismatch of $k$ between the gated and ungated regions increases, which results in the observed increase in reflection from the interface between gated and ungated regions. Once $V_g$ reaches $V_{th}$, the 2DES channel is fully pinched off, so the amplitude of the reflection cannot increase any further.

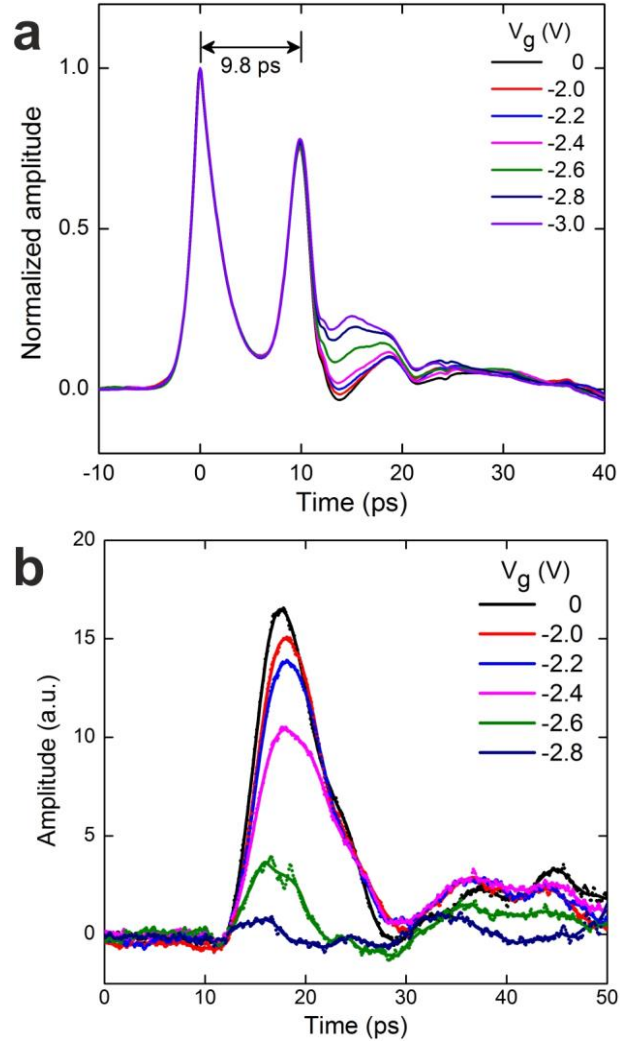

**Figure 2. Input and transmitted THz pulse through a 2DES channel.** (a) The measured input (and reflected) pulses for different $V_g$ at 4 K. The THz pulse is generated by S1 and detected by S2. (b) The measured transmitted signals through the 2DES for different $V_g$ at 4 K. The coplanar mode THz pulse is generated by biasing both S3 and S4, with detection by S2.

In the following, the coplanar mode THz pulse was excited by positively biasing a pair of PC switches (S1 and S2) which are located at either side of the CPW centre conductor, and the transmitted pulses at different $V_g$ were measured by a switch (S3) at the other side (see Supplementary Note 3).[12,20] We found that, by decreasing $V_g$ from 0 V to $V_{th}$, the transmitted coplanar mode signal showed a strong dependence on $V_g$. When the 2DES channel is pinched-off (*i.e.* for $V_g \leq -3$ V), there is still a sizable signal, caused by a crosstalk signal resulting from the capacitive coupling of pulses by between the ohmic contacts.[21] This parasitic signal does not depend on $V_g$, however, allowing us to obtain the pulse transmitted through the 2DES channel by subtraction of the crosstalk signal for a pinched-off channel (at $V_g = -3$ V, T = 4 K). As shown in Figure 2b, the obtained transmitted signal does not alter significantly between 0 V and $-2$ V. However, as $V_g$ approaches $V_{th}$, the mismatch in $k$ between the gated and ungated regions greatly increases, owing to the decrease of $n_s$ and $v_p$ in the gated region (Supplementary Figure S4), leading to a sharp decrease in transmission. In addition, the full-width-half-maximum (FWHM) of the transmitted pulse at $V_g = 0$ V is approximately 9 ps, which is much larger than that of the input pulse width (~1.5 ps). The contribution to this pulse broadening within the CPW region is 1.2 ps, judged by comparing the FWHM of reflected pulse (~2.7 ps) which propagates the same distance within CPW as the transmitted pulse (Figure 2a), with this input pulse width. Therefore, the observed pulse broadening (of ~9 – 1.5 ps = 7.5 ps) is mainly caused by the strong dispersion of plasmons in the 2DES, and especially in the ungated regions where their dispersion is greater.[17,22]

**Electrostatic modulation of 2D plasmons**

The dynamics of 2D plasmons in the 2DES can be obtained by measurements of the transmitted pulse as a function of $V_g$. However, comparative data obtained by subtracting pulses measured at different times exhibited poor signal-to-noise ratio (SNR), owing to a very slow drift of the laser power and/or focus position. We therefore employed a different technique for more detailed measurements, whereby a small AC signal ($V_{mod}$) was superimposed onto the DC gate bias, allowing lock-in detection (see Figure 1a). The time-resolved current of the transmitted pulse is denoted by $I(t,V_g)$, and its change ($\Delta I(t,V_g)$) with the swing of gate bias around the average voltage $V_g$, i.e. $\Delta I(t,V_g)/\Delta V_g$, is extracted using standard lock-in techniques. If the amplitude of $V_{mod}$ is sufficiently small, the measured signal is equivalent to $dI(t,V_g)/dV_g$ (see Supplementary Note 5). By altering $V_g$ at a fixed $V_{mod}$, changes in the transmitted THz signal as a function of $V_g$ could then be recorded. This gate-modulation technique provided an improvement in SNR of more than 50 times (see Supplementary Note 6 and Figure S6).

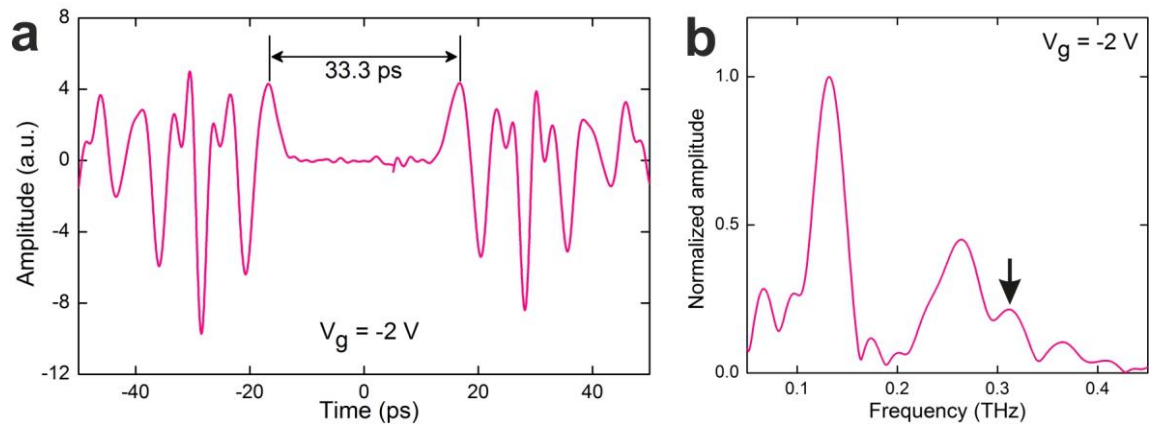

**Figure 3. Modulation of 2D plasmons in a 2DES channel.** (a) A superposition of the measured gate-modulation signals as a function of time for $V_g = -2.0$ V, in which the THz pulses propagate from S1 to S3 (positive time domain) and from S3 to S1 (negative time domain). (b) The FFT spectrum of the gate-modulation signal measured at $V_g = -2.0$ V. The grey regions are 80-GHz-wide frequency bands centred at the resonance frequencies of the fundamental and the second plasmon modes in the gated region. The resonance at $311 \pm 6$ GHz is indicated by a black arrow.

The detailed gate-modulation measurements were performed in a closed-cycle He$_3$/He$_4$ dilution refrigerator (see Methods for details).[12] Figure 3a shows the gate-modulation signals obtained at $V_g = -2$ V, when a PC switch bias ($V_{DC}$) of +5 V was applied to S1 to generate a picosecond pulsed signal propagating towards S3. In order to remove high-frequency noise in the data, without losing useful information, the measured signals were passed through a digital low-pass filter with a cut-off frequency of 0.6 THz, chosen to be higher than the upper frequency of the gate-modulated signals (~0.4 THz). The time taken for the pulse to propagate through the structure corresponds to the sum of the transit time in the CPW ($t_{CPW}$), and that in the 2DES ($t_{2DES}$). Measurement of pulses generated at S1 and measured at S3 were superposed with a measurement of the reverse-transmitted THz signal (*i.e.* propagating from S3, biased at +5 V, to S1) revealing a total round-trip time, $\Delta t = 2(t_{CPW} + t_{2DES})$, of 33.3 ps (Figure 3a). The measured gate-modulation signals were found to be symmetric for the two propagation directions (which was verified by swapping the excitation and detection switches). The $t_{CPW}$ was found to be ~9.8 ps, by measuring the time delay between the input pulse and its reflection from CPW/2DES interface (Figure 2a). Thus, the $t_{2DES}$ is ~6.9 ps. The corresponding average pulse velocity in the 2DES is ~1.1×$10^7$ m/s, which is an order of magnitude smaller than that in the CPW (~1.1×$10^8$ m/s) and close to the expected plasmon velocity in 2DES, supporting our assumption of plasmonic excitation. In addition, the transmitted signals shown in Figure 3a are composed of clear periodic oscillations. When a picosecond pulse is injected into the 2DES, the frequency components of the pulse that satisfy the Fabry-Perot resonance conditions of the gated or ungated plasmonic cavities are trapped, and undergo leaky oscillations within the cavity. Since $n_s$ in the gated cavity changes is tuned by $V_g$, the resonant signals from the gated plasmonic cavity are strongly modulated by the DC component of $V_g$. Therefore, the dynamic process of the resonant excitation and decay of plasmonic waves in the gated cavity is recorded in the time-domain gate-modulation signals. In the corresponding fast Fourier transform (FFT) spectrum (Figure 3b), the two strongest plasmon resonance frequencies observed are $132 \pm 2$ GHz and $264 \pm 2$ GHz, corresponding to the fundamental and second plasmonic mode of the gated plasmonic cavity, respectively. Using $n_s = 4.9×10^{15}$ m$^{-2}$ at the same $V_g = -2$ V, obtained from the *in situ* two-terminal magnetotransport measurement (see Supplementary Note 4), the predicted plasmon resonance frequencies in the gated region calculated using Equation (2) are 136 GHz ($n = 1$) and 264 GHz ($n = 2$), which are close to our measured values. There is also a resonance mode around 311 GHz in Figures 3b (indicated by black arrow), which can be attributed to the fundamental plasmon mode in the 19.7-µm-long ungated region (the calculated frequency using Equation (1) is 306 GHz). Unlike the modes formed by the gated cavity, this resonance does not change frequency when $V_g$ is altered. Aside from these three main resonance modes, there are several other weaker resonance peaks in the whole spectrum, which can be classified as coupled modes of two neighbouring cavities for low frequency (<100 GHz) resonances, and higher order modes.

To investigate the dependence of the plasmon resonances on $V_g$, we measured the time-domain profiles of the gate-modulation signals as a function of $V_g$ (Figure 4a). As $V_g$ is swept toward $V_{th}$, $n_s$ in the gated region decreases, and correspondingly the plasmon resonance frequency in the gated region is reduced. This results in an increase in the observed plasmonic oscillation period of the time domain signals. In the corresponding FFT spectra (Figure 4b), the frequency of the first gated plasmon mode is tuned from 158 GHz to 110 GHz by sweeping $V_g$ from 0 V to –2.5 V. The frequency of the second mode experiences a red-shift approximately twice that of the first mode with decreasing $V_g$. From the measured frequencies of the gated plasmon modes in Figure 4b, we calculated $n_s$ for each value of $V_g$ using Equation (2) and compared them with the measured $n_s$ from magnetotransport measurements, as shown in Figure 4c, and found good agreement between these values.

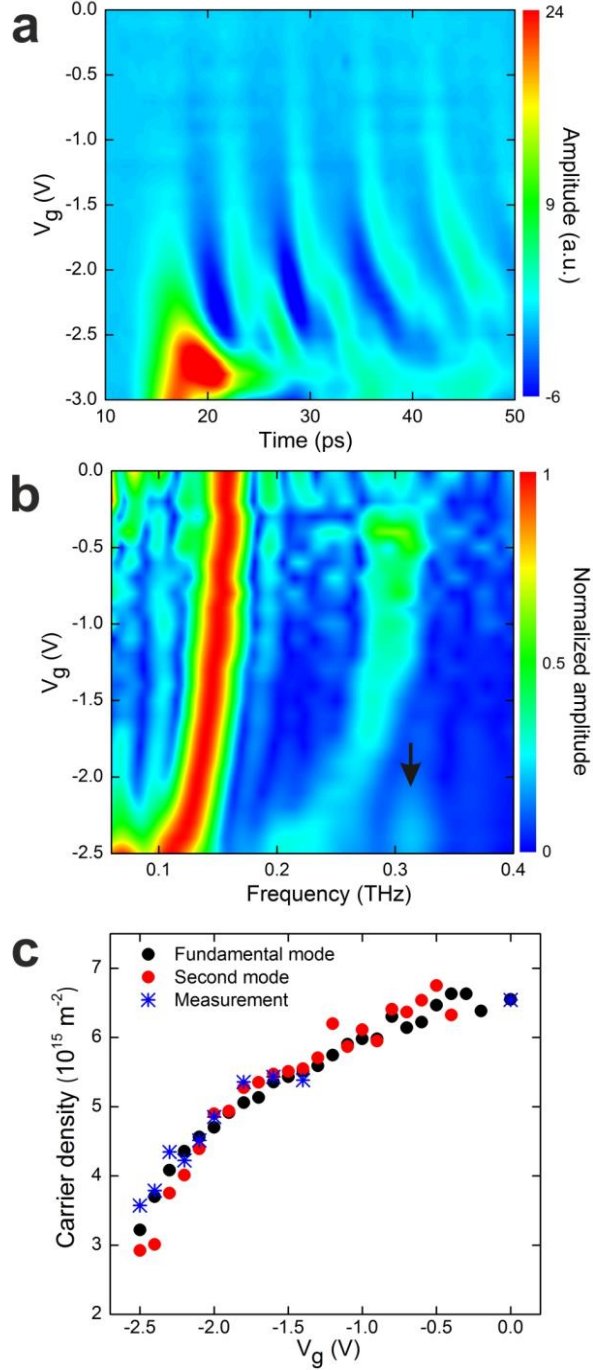

**Figure 4. Electrostatic manipulation of 2D plasmon resonances.** (a) Colour-scale plot of the measured gate-modulation signals plotted as a function of time and $V_g$. (b) Colour-scale plot of the normalized FFT spectra for gate-modulation signals measured as a function of frequency and $V_g$. The resonance around 311 GHz for $V_g \lesssim -2.0$ V is indicated by a black arrow. (c) The $n_s$ in the gated region as a function of $V_g$. These are calculated using the measured frequencies of fundamental (black dots) and second (red dots) plasmon modes and from the two-terminal magnetotransport measurements of the same 2DES mesa (blue asterisks).

## Conclusion

We have demonstrated an on-chip THz system that allows the picosecond-duration excitation, time-resolved detection and electrostatic manipulation of 2D plasmons. Using a gate-modulation technique, we observed the picosecond-resolved time evolution of confined 2D plasmons. This work not only provides a useful technique with which to study the ultrafast THz response and carrier dynamics of low-dimensional semiconductor structures, but also has potential future application in the development of more complex plasmonic circuits capable of manipulating THz waves. We note that in the system studied here, the properties of the 2DES could in principle be explored at even higher frequencies if the bandwidth of the plasmonic circuit could be expanded, with substrate thinning of the integrated wafer being one potential route to achieve this goal.[23]

## Methods

### Device fabrication

The wafer from which the device was fabricated was grown by MBE with the layer structure shown in Figure 1b. To obtain $n_s$ and $\mu$ of the 2DES, a Hall bar was fabricated, and magnetotransport measurements were undertaken in a 1.2 K helium bath cryostat with a superconducting magnet. The $n_s$ and $\mu$ of the 2DES in dark state were $3.66 \times 10^{15}$ m$^{-2}$ and 48.7 m$^2$/(V·s), respectively. After illumination by a HeNe laser, a $n_s$ of $6.26 \times 10^{15}$ m$^{-2}$ and $\mu$ of 89.5 m$^2$/(V·s) were obtained. The corresponding momentum relaxation time ($\tau_p$) was calculated to be 33 ps.

To fabricate the device, firstly a 100×25 µm$^2$ mesa containing the 2DES was defined using a sulphuric acid etch (H$_2$SO$_4$:H$_2$O$_2$:H$_2$O=1:8:70) to a depth of 100 nm. Next, two 12.5×30 µm$^2$ ohmic contacts to the 2DES were formed by depositing Au-Ge-Ni alloy (200 nm) at either end of the 2DES region, followed by annealing at 430 ℃ for 80 seconds. A continuous 4.4 µm-wide gate was then defined on top of the mesa using electron-beam lithography, and subsequent deposition of Ti/Au (10/60 nm) by electron-beam evaporation. To expose the underlying PC LT-GaAs layer, the region immediately surrounding the 2DES was protected using photoresist (Shipley S1813), and the exposed surface then etched in a selective etchant mixed from 50% citric acid and 30% H$_2$O$_2$ at the ratio of 3:1, down to an AlAs etch-stop layer, which was removed subsequently in dilute HF (5% concentration), to reveal a smooth LT-GaAs surface. A CPW structure, designed to incorporate two pairs of symmetric PC switches, was next defined by photolithography, and metallised via electron-beam evaporation of Ti/Au (10/150 nm). In the CPW structure, the widths of centre conductor and gaps were 30 µm and 20 µm, respectively. CPW centre conductor was aligned to the ohmic contacts formed previously on the 2DES. Corresponding breaks in the ground plane were incorporated, both to suppress crosstalk, and to allow contact to be made to the 2DES gate.

### Cryogenic measurement in 4K cryostat and dilution refrigerator

For measurements performed at 4 K, the sample was mounted in a continuous-flow liquid helium cryostat, with optical access provided through a z-cut quartz window.[19] The laser beams were focused through the quartz window onto the PC switches. The average laser power of both the pump and probe beams was fixed at 10 mW.

For measurements below 4 K, and in magnetic fields, the sample was mounted on a holder attached to mixing plate of a closed-cycle $He^3/He^4$ dilution refrigerator, located at the centre of a superconducting magnet. The experimental setup comprised in-situ piezoelectric stages used to position the laser beams on the sample surface, as discussed in detail in Ref. 12. Both the pump and probe laser beam average power were set to be 2 mW, which resulted in heating of the sample stage to around 2 K, measured using an embedded $RuO_2$ thermometer immediately adjacent to the device.

**Experimental setup for input and transmitted pulse measurements**

For input pulse measurements (see Supplementary Figure S2a), shunt interconnects were used to apply a DC bias to a PC switch (S1) which, when illuminated using a 100 fs pulsed, 800 nm near infrared (NIR) laser, generated a THz pulse that was coupled into the overlaid CPW structure. The unbiased PC switch immediately adjacent to S1 (S2) was illuminated by a time-delayed and mechanically chopped NIR laser beam, allowing time-resolved detection of the THz pulse launched into the CPW, and of reflected pulses generated from the 2DES. The current flowing in S2 was then extracted using standard lock-in techniques, with an optical chopper used to provide reference.

For transmitted pulse measurements (see Supplementary Figure S3a), a pair of PC switches (S3 and S4) were both biased positively with respect to the centre conductor and illuminated by a defocused pump laser beam, in order to generate a coplanar mode within the waveguide[20]. The PC switch on the opposite side of 2DES mesa (S1), illuminated by the chopped probe beam, was then used to detect pulses which had propagated along the CPW and through the attached 2DES. In order to extract measurements of pulses transmitted through 2DES mesa, the crosstalk signal between two ends of CPW was first measured by pinching off the 2DES channel ($V_g = -3$ V). The signals transmitted at different $V_g$ were measured in the following. By subtracting the crosstalk signal from each measured signal, the pulses transmitted through 2DES were obtained.

**Experimental setup for gate-modulation measurement**

As shown in Figure 1a, S1 and S3 were illuminated by the pump and probe laser beams, respectively. The THz pulse was generated by S1, which was biased by a DC voltage. Both $V_g$ and $V_{mod}$ were applied on the metallic gate simultaneously to control $n_s$ underneath the gated region. $V_g$ and $V_{mod}$ (the latter an 87 Hz sinusoidal wave, with 25~100 mV rms amplitude) were supplied by a digital-to-analog converter and an ultra-pure sinewave oscillator, respectively. The gate-modulation signal was measured at S3 using a lock-in amplifier, whose reference signal was provided via the synchronous output of the sinewave oscillator. To facilitate the alignment of laser beam in dilution refrigerator, we use one rather than two PC switches to generate THz pulse. Though the pulse generated from a single PC switch is a mixture of coplanar and slotline modes, only the coplanar mode signal could pass through the 2DES and be modulated by $V_{mod}$ (see Supplementary Note 3). In this configuration, the THz pulse propagates through the 2DES in the direction from S1 to S3. To measure the gate-modulation signal for a THz pulse propagating in the opposite direction, the electrical connections of S1 and S3 were swapped whilst maintaining fixed laser beam positions, and an equivalent measurement was taken. In this case, the laser beams swap functionality (*i.e.* the pump beam is now the probe beam, and *vice versa*).

## Acknowledgements

We gratefully acknowledge funding from EPSRC, the ERC ("NOTES" and "TOSCA" projects), the Royal Society, and the Wolfson Foundation. We are also grateful to Oleksiy Sydoruk and Andrey V. Shytov for useful discussions.

## Author contributions statement

J.B.W. and C.D.W. designed the device. J.B.W., W.M., M.C.R. and L.C. fabricated the device. L.H.L. grew the integrated wafer. C.D.W. and D.M. assembled the experiment. J.B.W., A.S.M., C.D.W. and D.M. performed the measurement and analysed the data. J.B.W., A.S.M., C.D.W. and J.E.C. wrote the manuscript. J.E.C. conceived and organized the project. All authors discussed the results and commented on the manuscript.

## Additional Information

**Supplementary information** accompanies this paper

**Competing financial interests:** The authors declare no competing financial interests.

# Supplementary Information

## Supplementary Note 1: Two-terminal conductance measurements of 2DES.

When the sample was cooled to 2 K, we found that switching on the laser resulted in an increase to the conductance through scattering of light onto the sample, and that this increase remained even after the laser was switched off, owing to persistent conductivity effects [1]. To measure the 2DES two-terminal conductance after illumination, an AC voltage (100 µV rms, 83 Hz) was applied across the 2DES and centre conductor of the CPW, with a DC bias ($V_g$) applied to the metallic gate, and lock-in detection was used to measure the conductance. The combined DC conductance of the CPW centre conductor and 2DES under illumination at 2 K is plotted in Supplementary Figure S1. The pinch-off voltage of the 2DES ($V_{th}$), *i.e.* the voltage for which the carriers beneath the gated region are completely depleted, was found to be −3.0 V.

## Supplementary Note 2: $V_g$ dependence of the input and reflected THz pulses.

The experimental arrangement used for the input and reflected pulse measurements is shown in Supplementary Figure S2a. The sample was mounted in a continuous-flow liquid helium cryostat, with the THz pulse generated at switch S1 and detected at S2. The average pump and probe beam powers were 10 mW, and $V_{DC}$ was set to be +20 V

The input and reflected pulses as a function of $V_g$ at 4 K were measured and each measured signal was normalized by the value of its first peak (input pulse) at 0 ps (Figure 2a of the main article). When $V_g$ is −3 V, the 2DES channel is pinched off, and the injected signal into the 2DES is completely reflected from the gated region. After subtracting the signal for $V_g = -3$ V from each signal for different $V_g$, we extract the relative change of reflected signals from the gated region. As shown in Supplementary Figure S2b, there is almost no signal at the peak position of input pulse (0 ps) and reflected pulse from ohmic contact (9.8 ps), indicating that these signals do not depend on $V_g$. Conversely, a strong peak appears at the 'shoulder' in Figure 2a of the article, which increases monotonically in amplitude as $V_g$ decreases. As mentioned in the main article, the increased reflection of the broadband THz signal with decreasing $V_g$ is due to the increased mismatch of wave number between gated and ungated region.

In Figure 2a of the main article, the delay time between the first and second peaks is 9.8 ps, which corresponds to the round-trip time of THz pulse in length of CPW connecting S1 and S2 to the 2DES mesa. The length of this CPW is 550 µm, from which the THz pulse velocity is calculated to be ~1.1×$10^8$ m/s in the CPW. Considering that the CPWs on the two sides of 2DES mesa are symmetric, we can therefore state that the single-trip transit time of a pulse propagating in the CPW between S1/S2 and S3/S4 without the 2DES mesa is 9.8 ps.

## Supplementary Note 3: Transmitted coplanar mode and slotline mode signals.

In the CPW, two main modes are supported: the coplanar mode and the slotline mode. There is significant difference in the field distribution for these two modes: for the coplanar mode, the electric field distribution is symmetric with respect to the CPW centre conductor, and the electric field lines lie between centre conductor and the two adjacent ground planes, whilst for the slotline mode propagation, the electric field lines lie between the two ground planes, bypassing the centre conductor. The radiative losses of the slotline mode are much greater than of the coplanar mode [2,3].

A diagram of selective mode generation and transmitted pulse detection is shown in Supplementary Figure S3a. A defocused pump beam is used to illuminate the two PC switches on the same side of the 2DES (*e.g.* S3 and S4). When the two switches are biased with the same positive voltage ($V_{DC1} = V_{DC2}$), a pure coplanar mode is generated and propagates in the CPW. For slotline mode generation, the polarity of one PC switch is altered, such that $V_{DC1} = -V_{DC2}$. The PC switch on the opposite side of the 2DES mesa and on either side of the CPW (*e.g.* S2 in Supplementary Figure S3a), which is illuminated by chopped probe beam, is used to detect the transmitted pulse in both cases. Under coplanar mode excitation, the transmitted pulses measured on both sides of the CPW (S1/S2) have the same pulse shape and polarity because of the symmetric field pattern of coplanar mode. When the slotline mode is excited, the pulse shape measured on either side of CPW is nearly the same, but the polarity is opposite, indicating that the field pattern of slotline mode is asymmetric.

The transmitted coplanar mode and slotline mode signals for different $V_g$ at 4 K are shown in Supplementary Figure S3b,c. The measurements were taken in a continuous-flow liquid helium cryostat. The average pump and probe beam powers were maintained at 10 mW. The biases on PC switches of S3 and S4 were: $V_{DC1} = V_{DC2} = +20$ V for coplanar mode excitation, and $V_{DC1} = -V_{DC2} = +20$ V for slotline mode excitation. The transmitted coplanar mode signal exhibits an obvious dependence on $V_g$, with decreasing amplitude as $V_g$ is varied from 0 V to −3 V (Supplementary Figure S3b). This indicates that the THz pulse is interacting strongly with the 2DES, and the coplanar mode transmission can therefore be controlled electrostatically by altering $V_g$. When the channel is pinched off ($V_g = -3$ V), a strong, broad transmitted pulse is still observed, which is attributed to crosstalk between the two ends of the 2DES mesa. By subtracting the transmitted pulses obtained for different $V_g$ from the crosstalk signal measured at $V_g = -3$ V, the transmitted pulses through 2DES mesa were obtained as shown in Figure 2b of the main article. In contrast, the transmitted slotline mode signals is almost unaffected as $V_g$ decreases from 0 V to −3 V as shown in Supplementary Figure S3c, indicating that the slotline mode THz pulse does not interact strongly with the 2DES. For the slotline mode propagating in the CPW, the electric field close to the centre conductor is much weaker than that of coplanar mode, and therefore the conversion efficiency from the slotline mode to 2D plasmons in the 2DES is low.

**Supplementary Note 4: Two-terminal magnetoresistance measurements and carrier density in the 2DES.**

For two-terminal magnetotransport measurements, magnetic field was applied normal to the 2DES plane and swept from 0 to 2.5 T. Supplementary Figure S4a shows the two-terminal resistance of the 2DES measured as a function of magnetic field at $V_g = -2$ V. The resistance of the leads and centre conductor of CPW were not subtracted. The resistance increases with magnetic field, and quantum Hall plateaux appear at high magnetic fields. For two-terminal measurements of a 2DES, the resistance in magnetic field depends on both diagonal and off-diagonal components of the resistivity tensor, and on the sample geometry. Therefore, the Shubnikov-de Hass (SdH) oscillations are not clearly resolved in Supplementary Figure S4a. By calculating the second derivative of resistance with respect to magnetic field ($d^2R/dB^2$), however, two oscillations of different periods are observed (Supplementary Figure S4b), corresponding to the SdH oscillations in the gated and ungated regions [4]. The periods of the SdH oscillations ($\Delta(1/B)$) were obtained from the FFT spectrum of $d^2R/dB^2$. As shown in Supplementary Figure S4c, the two peak frequencies correspond to the periods of 0.100 T$^{-1}$ and 0.074 T$^{-1}$. Using the relationship: $\Delta(1/B) = 2e/n_sh$, where $e$ is the elementary charge and $h$ is Planck's constant, the electron concentrations, $n_s$ are calculated to be 4.9×10$^{15}$ m$^{-2}$ and 6.5 ×10$^{15}$ m$^{-2}$, respectively.

By measuring the resistance as a function of magnetic field at different $V_g$, the $V_g$ dependence of $n_s$ was obtained. For $V_g$ = 0 V, only one SdH oscillation period ($\Delta(1/B)$ = 0.074 T$^{-1}$) was observed, with a corresponding $n_s$ of 6.5 ×10$^{15}$ m$^{-2}$, indicating that the gated and ungated regions have the same $n_s$ in this case. As $V_g$ is swept below −1.4 V, two peaks emerge in the FFT spectrum of $d^2R/dB^2$. As $V_g$ decreases further, the peak at the higher frequency remains constant (as it arises from the ungated region) while the lower frequency peak gradually decreases in frequency (since it arises from the gated region). As shown in Supplementary Figure S4d, $n_s$ in the ungated region remains approximately constant at a value of ~6.5×10$^{15}$ m$^{-2}$, while the $n_s$ in the gated region gradually decreases from 6.5×10$^{15}$ m$^{-2}$ for $V_g$ = 0 V, to 3.6×10$^{15}$ m$^{-2}$ for $V_g$ = −2.5 V.

## Supplementary Note 5: Dependence of gate-modulation signals on the amplitude of $V_{mod}$ and temperature.

To select the optimum value of $V_{mod}$ for measurements, gate-modulation signals with different amplitudes of $V_{mod}$ were compared. Measurements were taken at a constant temperature of 2 K using the same experimental arrangement shown in Figure 1a of the main article. $V_{DC}$ was set to be +5 V, and the average power of both pump and probe beam were 2 mW. When $V_g$ is around −2 V, the amplitude of the plasmonic oscillation is the strongest (Figure 4a of the main article), and so this point was selected for the optimisation.

During data processing, the measured gate-modulation signals for different $V_{mod}$ were divided by the corresponding amplitude $V_{mod}$ for normalization and filtered by a low-pass filter with a cut-off frequency of 0.6 THz to remove high-frequency noise. The gate-modulation signals for different $V_{mod}$ (10, 25, 50 and 100 mV rms) are shown in Supplementary Figure S5a. When the amplitude of $V_{mod}$ was increased from 10 to 100 mV rms, the two periodic oscillations are still apparent, but the oscillations are observed to decay faster for $V_{mod}$ = 100 mV rms than at lower values of $V_{mod}$. In the corresponding frequency spectra, the first and second gated plasmon modes are all discernible and are almost constant (Supplementary Figure S5b). However, the plasmonic resonances for $V_{mod}$ = 100 mV rms are not as sharp as the signals for smaller $V_{mod}$, in accordance with the larger decay rate of oscillations in the time domain.

The difference in decay rates for different $V_{mod}$ can be explained by a change in quality ($Q$) factor of the gated plasmonic cavity. In the gated plasmonic cavity under application of a DC bias, the $Q$ factor $Q = f_r/\Delta f_0$, where $f_r$ is the resonance frequency and $\Delta f_0$ is the half power bandwidth. Here, $\Delta f_0$ is determined by the losses in the Fabry-Perot cavity, principally ohmic losses in the cavity and coupling losses at the cavity boundaries. When $V_{mod}$ is applied in addition to $V_g$, the plasmonic resonance frequencies are tuned dynamically around the centre frequency $f_r$, and the tuning range is denoted by $\Delta f_{mod}$. The resonance bandwidth of the gated plasmonic cavity is then widened by $\Delta f_{mod}$. In this case, the $Q$ factor of the gated plasmonic cavity, which is being modulated by $V_{mod}$, can be given by $Q = f_r/(\Delta f_0 + \Delta f_{mod})$. Based on Equation (1) of the main article, we obtain

$$\mathrm{d}f \propto \frac{f}{n_s} \cdot \mathrm{d}n_s, \tag{s1}$$

where $f$ is the gated plasmon frequency. For constant $V_g$, the change of $n_s$ ($\Delta n_s$) increases monotonically with the amplitude of $V_{mod}$. Therefore, the increase of amplitude of $V_{mod}$ results in an increase of $\Delta f_{mod}$, and the $Q$ factor correspondingly decreases. As a result, plasmonic oscillations for $V_{mod} = 100$ mV rms exhibit the largest decay rate observed in the time-domain.

Based on the above analysis, a trade-off between the SNR and accuracy of the plasmon resonance frequency measurement is required in order to select $V_{mod}$ in gate-modulation measurements. For example, in our measurements, the amplitude of $V_{mod}$ was set to be between 25 and 100 mV rms. When the plasmonic oscillation was strong, e.g. when $V_g = -2$ V and magnetic field is 0 T, we chose $V_{mod} = 25$ mV rms in order to obtain precise plasmon resonance frequencies. As $V_g$ approaches 0 V, the plasmonic oscillation becomes relatively weak, so a large $V_{mod}$ is needed to improve SNR, and $V_{mod} = 100$ mV rms was usually used.

To study the temperature dependence, the gate-modulation signals as a function of temperature were measured. As shown in Supplementary Figure S5c, the gate-modulation signals for $V_g = -2$ V are very similar at the three measured temperatures of 2 K, 4.5 K, and 10 K, implying that the excitation and propagation of 2D plasmons in the 2DES mesa does not change significantly in the temperature range 2–10 K. In the corresponding frequency spectra (Supplementary Figure S5d), the plasmonic resonance frequencies of the first and second modes in the gated regions do not show obvious change as the temperature increases from 2 K to 10 K, indicating that $n_s$ does not change significantly in this temperature range.

**Supplementary Note 6: Signal-to-noise ratio (SNR) of the gate-modulation technique**

As discussed in the main article, the signal measured using the gate-modulation technique can be approximated to d$I(t)$/d$V_g$. Here, we compare the SNR of d$I(t)$/d$V_g$ for $V_g = -2.5$ V using using both the difference pulse, and the gate-modulation technique. For both methods, measurements were performed in a liquid-helium continuous flow cryostat at 4 K, with the same setup for coplanar mode excitation. For the difference pulse method, the measured transmitted coplanar mode pulses for $V_g = -2.4$ V and $-2.6$ V are shown in Supplementary Figure S3b. d$I(t)$/d$V_g$ for $V_g = -2.5$ V was obtained by the central difference: ($I(t)(V_g = -2.4$ V$) - I(t)(V_g = -2.6$ V$))/((-2.4$ V$) - (-2.6$ V$))$. For the gate-modulation technique, an AC voltage ($V_{mod}$) of amplitude 100 mV rms was superimposed onto $V_g$, whilst the optical chopper used to modulate the probe beam in the previous measurements was removed. In this case, the modulated signal for $V_g = -2.5$ V was measured at S2 using a lock-in amplifier with reference signal supplied from the AC voltage source.

Signals measured using the difference pulse method and gate-modulation technique are normalized and plotted in Supplementary Figure S6. The shape of the gate-modulation signal resembles closely that of the difference pulse signal, but a significant improvement in SNR is observed. Considering that the delay time for THz in CPW is 9.8 ps, the real signal measured at times earlier than 9.8 ps should be zero (Figure 2a of the main article). We can therefore use the time-domain signal fluctuation observed in the time window from 0 to 9.8 ps to study the noise levels of the two methods. The fluctuations in signal observed when using the difference pulse method are in excess of 10%. When using the gate-modulation technique, however, these fluctuations are reduced to only 0.2%, so demonstrating a 50 times improvement in SNR

**Supplementary Figures**

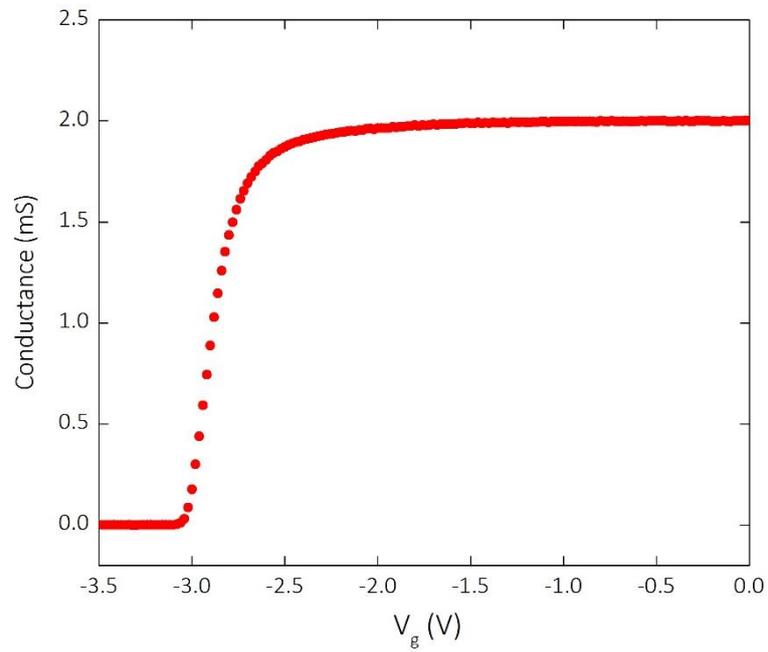

**Supplementary Figure S1. Measured conductance of the 2DES mesa as a function of gate voltage $V_g$ at 2 K.** The resistance of the leads and centre conductor of the coplanar waveguide (CPW) are not subtracted.

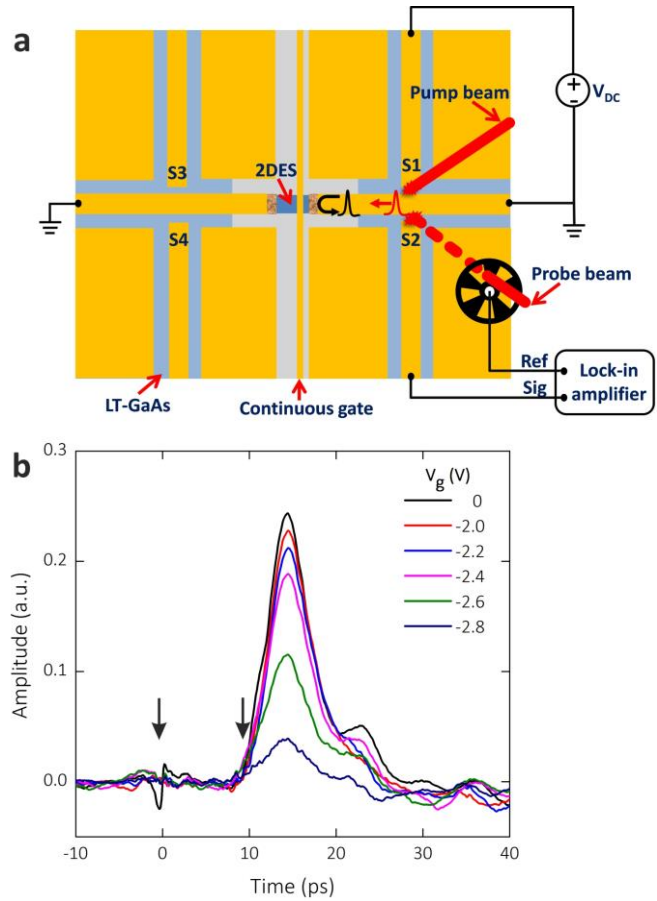

**Supplementary Figure S2. Input and reflected THz pulse measurement at 4 K.** (a) The experimental arrangement used to measure the input and reflected THz pulses. A THz pulse is generated at switch S1, which is illuminated by a near-infrared (NIR) pump laser beam and biased by a positive DC voltage ($V_{DC}$). Switch S2 is illuminated by an optically chopped and time-delayed probe laser beam, allowing detection of the pulse generated at S1 and the pulse reflected from the 2DES mesa. The current induced in S2 by the THz-frequency electric field is recorded by a lock-in amplifier, referenced to the optical chopper frequency. (b) The change in the reflected signal relative to that found at $V_g = -3$ V for different $V_g$ at 4 K, and obtained by subtracting the signal at $V_g = -3$ V in Figure 2a of the main article from each signal recorded at different $V_g$. The peak positions of input pulses (0 ps) and the reflected pulses from ohmic contact (9.8 ps) in Figure 2a of the main article are indicated using black arrows.

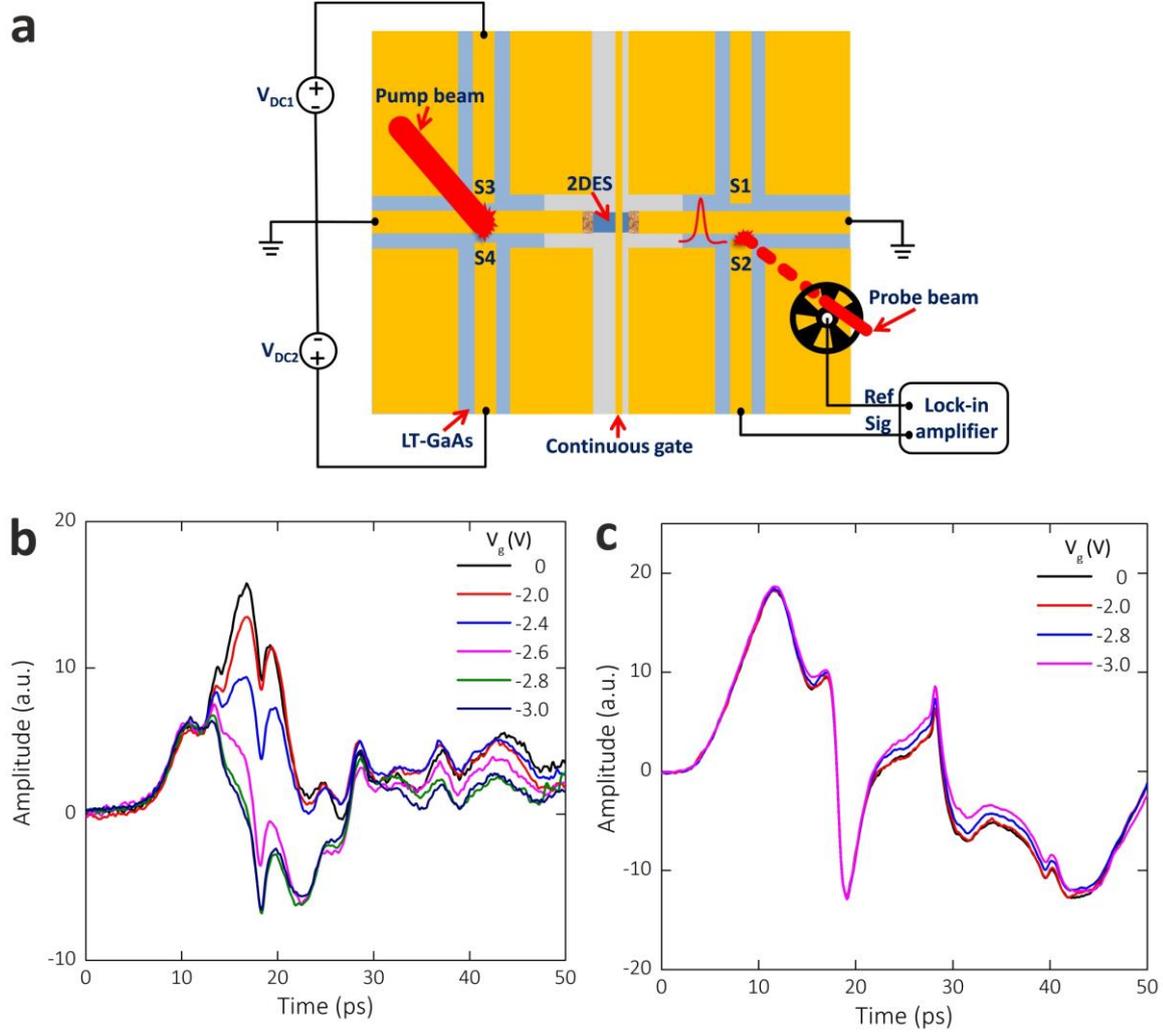

**Supplementary Figure S3. Dependence of the transmitted coplanar mode and slotline mode signals in the CPW on $V_g$.** (a) The experimental apparatus used for selective mode generation and transmitted signal measurements. Adjacent PC switches located on one side of the mesa (*e.g.* S3 and S4) are illuminated by a defocused pump laser beam, and are biased with $V_{DC1}$ and $V_{DC2}$ respectively to select a particular mode. To excite a coplanar mode, $V_{DC1} = V_{DC2}$, while to excite the slotline mode, $V_{DC1} = -V_{DC2}$. Transmitted pulses are detected by illuminating a third PC switch, located on the opposite side of 2DES mesa (*e.g.* S2), using a time-delayed and optically chopped probe beam. (b) The transmitted coplanar mode signals ($V_{DC1} = V_{DC2} = +20$ V) for different $V_g$ at 4 K. (c) The transmitted slotline mode signals ($V_{DC1} = -V_{DC2} = +20$ V) for different $V_g$ at 4 K.

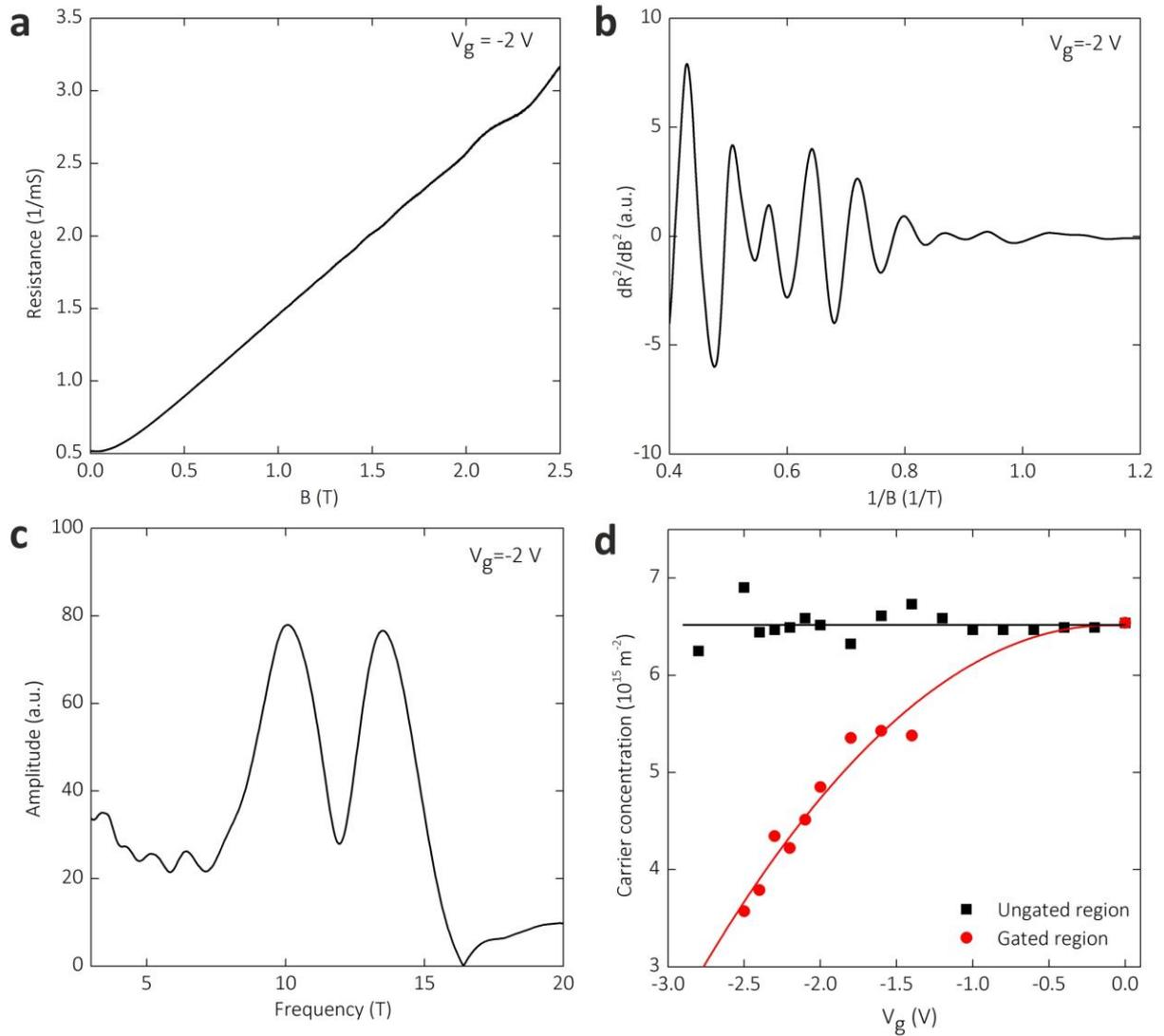

**Supplementary Figure S4. Two-terminal magnetotransport measurements and determination of the electron concentration ($n_s$) in the 2DES mesa.** (a) The two-terminal resistance of the 2DES mesa as a function of magnetic field ($B$), in which the resistance of the leads and the centre conductor of the CPW are not subtracted. (b) The second derivative of $R$ with respect to $B$ ($d^2R/dB^2$) as a function of $1/B$. (c) The fast Fourier transform (FFT) spectrum of $d^2R/dB^2$ shown in (b). (d) Calculated $n_s$ in the ungated (black squares) and gated (red circles) regions as a function of $V_g$. The solid lines are to guide the eye and indicate that $n_s$ in the ungated region (black) is constant while $n_s$ in the gated region (red) decreases when $V_g$ decreases from 0 V to –3 V, respectively.

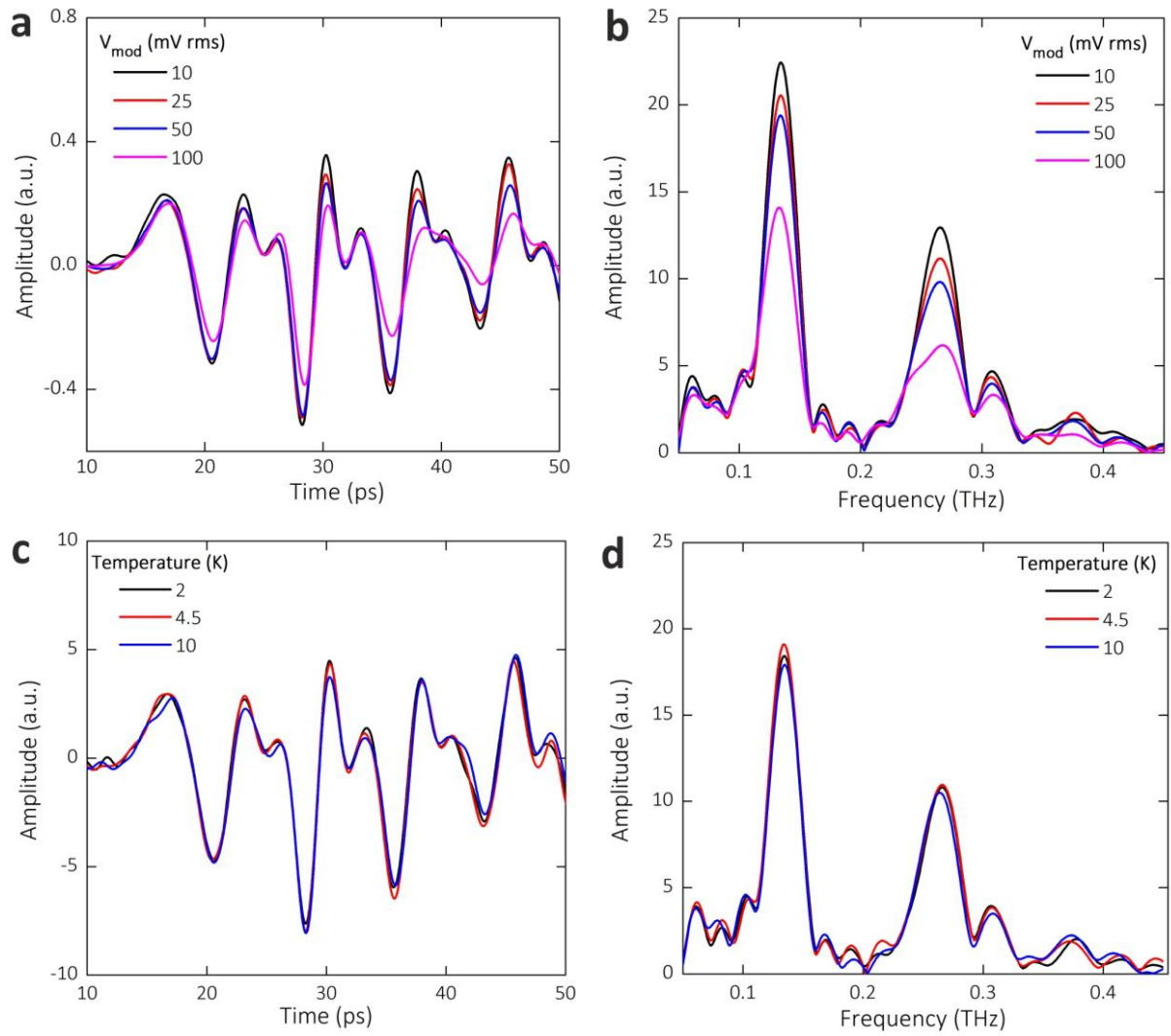

**Supplementary Figure S5. Dependence of the gate-modulation signals on** $V_{mod}$ **and temperature.** (a) The measured gate-modulation signals for $V_g = -2$ V at 2 K with different amplitudes of $V_{mod}$, ranging from 10 mV rms to 100 mV rms. Each signal is divided by the corresponding amplitude of $V_{mod}$. (b) The corresponding FFT spectra of the signals in (a). (c) The measured gate-modulation signals for $V_g = -2$ V at different temperatures. (d) The corresponding FFT spectra of the signals in (c).

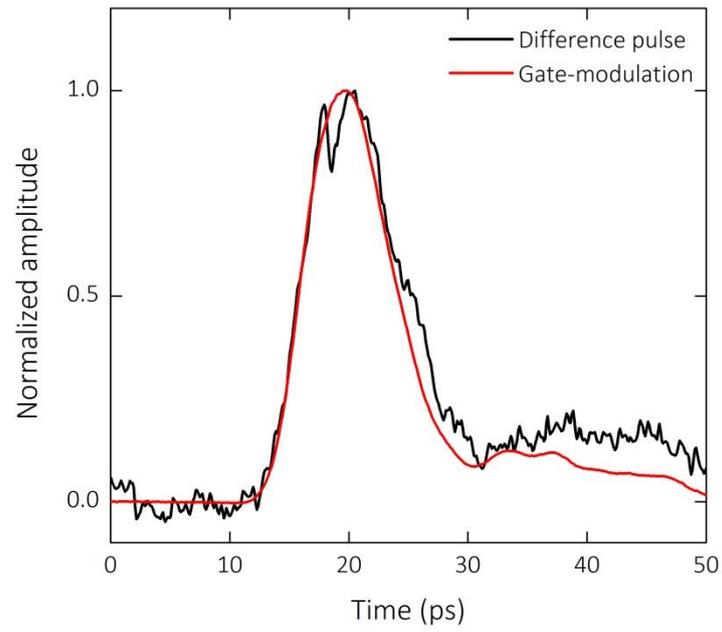

**Supplementary Figure S6. Comparison of the derivative of transmitted pulse with respect to** $V_g$ **(d$I(t)$/d$V_g$) for** $V_g$ **= –2.5 V using the difference pulse method and gate-modulation technique.**

## Supplementary References